\documentclass[aps,showpacs,twocolumn,floatfix,amsmath,preprintnumbers,nofootinbibe,secnumarabic]{revtex4} 

\usepackage{graphicx,amsmath,amssymb,hyperref,url}

\usepackage{color}

\usepackage[caption=false]{subfig}
\def\sc{0.33}  

\def\art{paper}

\def\boo#1#2#3#4#5#6#7{\textit{#2} (#3, #4, #6).} \def\jrn#1#2#3#4#5#6{{#3} \textbf{#4}, #5 (#6).} \def\andd{, and } \def\Ref{} \def\Refs{} \def\Fig{Fig. } \def\eq{Eq. }  \def\eqref#1{(\ref{#1})} \def\Sec{Sec. }  

\def\bfl{\begin{flushleft}}
\def\efl{\end{flushleft}}
\def\bfr{\begin{flushright}}
\def\efr{\end{flushright}}
\def\bc{\begin{center}}
\def\ec{\end{center}}
\def\be{\begin{equation}}
\def\ee{\end{equation}}
\def\bse{\begin{subequations}}
\def\ese{\end{subequations}}
\def\ba{\begin{eqnarray}}
\def\ea{\end{eqnarray}}
\def\baa#1{\begin{array}{#1}}
\def\eaa{\end{array}}
\def\bw{} \def\bw{\begin{widetext}}
\def\ew{} \def\ew{\end{widetext}}
\def\nn{\nonumber }
\def\lb#1{\label{#1}}
\def\bit{\begin{itemize}}
\def\eit{\end{itemize}}
\def\bco{}
\def\bcs{\begin{cases}}
\def\ecs{\end{cases}}

\def\Der#1#2{\frac{\drm #1}{\drm #2}}

\def\Sin#1#2{\, \text{sin}^{#1}#2}

\def\drm{d}

\def\dn{\rho}  \def\dnc{\bar{\dn}}
\def\lnc{\bar{\ell}}

\def\tfc{q}

\def\nc0{b_0}

\def\wfv{\Psi_{\text{vac}}}

\def\lpwr{L_\pwr}

\def\rcos{R_{\text{dS}}}

\def\ssmgl{{\virgo}}
\def\ssmgl{{\text{mgl}}}
\def\ssmgl{{M}}
\def\ssmgl{{\mathfrak M}}

\def\rad{r}

\def\pwr{\chi}

\def\ror{R}

\def\mesph{\sigma^2} 

\begin{document}






\preprint{\small \footnotesize Astron. Rep. \textbf{65}, 1078-1083 (2021)   
\ \ 
[\href{https://doi.org/10.1134/S1063772921100437}{DOI: 10.1134/S1063772921100437}]
}

\title{
On asymptotic behaviour of galactic rotation curves in superfluid vacuum theory\footnote{Paper 
presented at the Fourth Zeldovich meeting, an international conference in honor of Ya.B. Zeldovich held in Minsk, Belarus, on September 7--11, 2020. Published by the recommendation of the special editors: S.Ya. Kilin, R. Ruffini and G.V. Vereshchagin.}
}

\author{\bf 
\firstname{K.~G.}~\surname{Zloshchastiev}}%
\affiliation{\mbox{Institute of Systems Science, Durban University of Technology, 
Durban 4000, South Africa}\\
e-mail: kostiantynz@dut.ac.za, kostya@u.nus.edu}


\begin{abstract}
The logarithmic superfluid theory of physical vacuum predicts that gravity is an induced phenomenon,
which has a multiple-scale structure.
At astronomical scales, as the distance from a gravitating center increases,
gravitational potential and corresponding spacetime metric 
are dominated by a Newtonian (Schwarzschild) term, followed by a logarithmic term, 
finally by linear
and quadratic (de Sitter) terms.
Correspondingly, rotation curves
are predicted to be Keplerian in the inner regions of galaxies, mostly
flat in the outer regions, and non-flat in the utmost outer regions.
We compare theory's predictions with the furthest rotation curves data points available
for a number of galaxies:
using a two-parameter fit,
we perform a preliminary estimate which disregards the combined effect of gas and stellar disc,
but 
is relatively simple and uses minimal assumptions for galactic luminous matter.
The data strongly points out at the existence of a crossover transition from flat to non-flat
regimes at galactic outskirts.
\end{abstract}


\date{received: 26 November 2020}


\maketitle

%
According to the Newton's theory of gravity, 
rotation curves,
i.e., rotation velocities 
as functions of distances from a gravitating center,
of free-falling non-relativistic test particles must have a Keplerian form,
which means that
their velocity must be inversely proportional to the square root of distance.
However, numerous astronomical observations, of both stars and luminous gas in galaxies, 
show significant deviation from this behavior.
For most of galaxies, the data shows rotation curves are flat (FRC),
i.e., the tangent lines of rotation curves are approximately horizontal,
for a wide range of  
distances from a center \cite{rvt80}. 

There are currently two main approaches to address this problem. 
One approach is popular among particle physicists working
in the Standard Model (SM) theory
of strong and electroweak interactions and its extensions.
This approach
assumes that some kind of corpuscular matter exists,
referred to as dark matter (DM), which interacts with known SM particles via gravity, but otherwise very weakly, if at all. 
Such matter seems to be truly elusive: its particles must be abundant in galaxies, including our Milky Way;
yet they do not, to the best of our knowledge, affect Earth's particle experiments.

The other approach assumes that DM-attributed phenomena are not caused
by any unknown corpuscular matter; but that gravity alone alters its behaviour from 
being Newtonian-like at length scales which are very different from our 
habitual scale.
This trivially resolves the question of the absence of SM-DM interactions in Earth's experimental conditions.  
The scale dependence of gravity also explains the significantly non-Keplerian behaviour of galactic rotation curves,
while orbits of planets and satellites remain Keplerian to a high degree of precision.
This approach is also very popular:
in fact, it is hard to find a theory of modified gravity which does not 
explain the non-Keplerian phenomenon 
of galactic rotation curves. 
However, this approach is not without its drawbacks.
First, because
the number of mutually exclusive gravity theories is currently so large,
this approach's \textit{status quo} lacks universality and unification.
More importantly, 
most theories of modified gravity (that we know of) require a different set of parameters for each galaxy;
which raises the question of why 
every galaxy, or even every isolated supermassive body, 
should ``have'' its own theory of gravity.

Therefore, 
some criteria for a convincing theory of astronomical scale phenomena are necessary;
in addition to those of \textit{bona fide} unification,
universality
across all galaxies,
and minimal free parameters.
One requirement is that a candidate theory should predict hitherto unknown phenomena
at a galactic scale, or even beyond, which can be empirically verified.
Another expectation is 
that it should describe phenomena, which are attributed not only to dark matter,
but also to dark energy (DE).
These would include the accelerating expansion of the Universe, Hubble law
and the notable discrepancy between outcomes
of the different measurement methods of the Hubble constant: 
those based on  cosmic microwave background radiation 
(suggesting a value $67.4 \pm 0.5$ km/s/Mpc
for a Hubble constant);
versus other methods,
such as distance ladder measurements using
Cepheids ($74 \pm 1$ km/s/Mpc) and
red giants ($70 \pm 2$ km/s/Mpc),
and geometric distance measurements
using megamasers ($74 \pm 3$ km/s/Mpc).

There is also growing consensus that a convincing explanation of DM- and DE-attributed phenomena cannot 
be based on a stand-alone model, but must
be a part of a fundamental theory, involving other fundamental interactions known to date.
Such a theory should definitely operate at an utterly quantum level, 
which implies reformulating the concept of gravity as a quantum phenomenon, commonly referred to as quantum gravity.

One of the most viable candidates for such a theory of physical vacuum and quantum gravity is superfluid vacuum theory (SVT),
a post-relativistic approach to high-energy physics and gravity.
This is essentially a framework,
which evolved from Dirac's idea of 
non-removable quantum matter filling an otherwise empty three-dimensional space \cite{dir51};
a pedagogical introduction can be found in monographs \cite{volbook,huabook}.
As a matter of fact,
superfluid vacuum theory possesses features of both corpuscular DM and modified-gravity approaches:
it is not only an essentially quantum many-particle theory in its foundations, 
but it also induces and includes relativistic gravity and spacetime
in the ``phononic'' or linear low-momentum limit of dispersion relations for excitations,
which are thus observed as relativistic particles.

Within the framework of the superfluid vacuum paradigm, various theories currently co-exist,
which share the main idea, but which offer different views of the dynamics and structure of the physical vacuum.
One of these theories is logarithmic superfluid vacuum theory,
whose foundations can be found in \Refs \cite{z10gc,z11appb}. 
To the best of our knowledge,
this model is not only free from the above-mentioned limitations, but also explains the
known DM- and DE-attributed phenomena, and
predicts a number of
new effects, including vacuum Cherenkov radiation, deformed dispersion relations, mass generation 
and superfluid stars \cite{z11pla,dz11,z19ijmpb,z20ijmpa,z20un1,z21ltp}. 
More relevant to the purposes of this \art, works \cite{z11appb,z20un1} describe a number of astronomical-scale effects predicted by logarithmic SVT;
including various expansion mechanisms and non-Keplerian (flat followed by non-flat) behaviour of galactic rotation curves.
The latter will be our main focus here.


%
Logarithmic superfluid vacuum theory assumes that 
physical vacuum is  quantum Bose liquid
described by a condensate wavefunction 
obeying
a nonlinear wave equation with logarithmic nonlinearity,
whose parameters are:
the inertial 
mass $m$ of fundamental background condensate's particles,
the critical density $\dnc$,
and 
the real-valued constant parameters $b_0$ and $\tfc$ of the nonlinear coupling.
A vast amount of mathematical
literature dealing with such equations exists, 
to mention just very recent examples \cite{zh20,aj20,ss20,tb20je,la20,zy20,yb21}.
In the general theory of superfluids,
logarithmic nonlinearity describes many-body effects -- 
in fact,
it can be used as a leading-order or robust approximation for a large class of strongly interacting
liquids in which the characteristic kinetic energies of particles
are less than the inter-particle potentials \cite{z18zna}.    
Logarithmic fluid models are known to work very well for laboratory superfluids,
such as the helium II phase, where they have resolved a number 
of long-standing problems \cite{z12eb,sz19} and predicted new effects \cite{z19ijmpb}.

When applying them to the theory of physical
vacuum, one can show that this background superfluid induces a four-dimensional spacetime,
while
photon-like excitations are somewhat analogous to  
acoustic waves in laboratory superfluids, traveling at the speed
$ c_b 
=
\sqrt{b_0/m}$ in the approximation of a homogeneous isothermal liquid \cite{z11appb,z20un1}.
In the ``phononic'' (low-momentum) limit,
$c_b
\to c_{(0)} \approx c $ 
where  
$c = 2.998 \times 10^{5}$ km/s
is 
called, for historical reasons, the speed of light in vacuum.

For our purposes here, we 
assume spherical symmetry, and
omit terms which decay faster than an inverse distance at spatial infinity, 
in which case the induced metric can be written 
in static coordinates
in the form  \cite{z20un1}: 
\ba
d s^2 
&=& 
- 
K^2 \,  c_{(0)}^2 d t^2
+ 
\frac{1}{K^{2}}  d r^2+ 
r^2 d \mesph
,
\lb{e:medsa2}\\
K^2  &\equiv&
1 
+ \delta_0 
- \frac{r_H}{r}
+ \beta_\pwr^2 
\ln{\!\left(
\frac{\rad}{\lnc}
\right)} 
+ \frac{r}{R_\ssmgl} - \frac{r^2}{\rcos^2}
, 
\nn
\ea
where
$d\, \mesph = d\theta^2 + \Sin{2}{\theta} \, d \varphi^2$
is the line element of a unit two-sphere,
$\delta_0 
=
2
(a_0 b_0 
+ a_2 \tfc/\lnc^2)/m c_{(0)}^2
$,
$\lnc = (m/\dnc)^{1/3}$, 
$r_H = 2 a_1 \tfc/ m c_{(0)}^2 \lnc$,
$ 
\beta_\pwr
=
\sqrt{
{2 \pwr \, b_0}/{m c_{(0)}^2}
}
$, 
$
R_\ssmgl
=
m c_{(0)}^2 \lnc/2 a_1 b_0 
$,
$\rcos 
= 
 \lnc
\sqrt{m  c_{(0)}^2/2 a_2 b_0}
$. 

One can see that the induced spacetime metric depends on a number of parameters, which can be divided into two groups. 
The first consists of parameters of the model's Hamiltonian: 
fundamental mass scale $m$, 
fundamental density scale $\dnc$,
and parameters of the nonlinear coupling, $b_0$ and $\tfc$.
However,
only two of these parameters, $m$ and $\dnc$,
are  \textit{ab initio} fixed; whereas the other two, $b_0$ and $\tfc$,
can vary depending on  
the environment, because 
the nonlinear coupling 
is a linear function of quantum temperature. 
The latter is defined as 
a thermodynamic conjugate
of quantum information entropy, sometimes dubbed as the Everett-Hirschman entropy,
and is conjectured to be linearly related to conventional (thermal) temperature \cite{z18zna}.

The second group are 
parameters $a$'s and $\pwr$ of a trial wavefunction of the state $|\wfv \rangle$ our superfluid is in,
which is defined as a solution of a logarithmic quantum wave equation, 
further details can be found in section 3 of \Ref \cite{z20un1}.
We expect that this state is stable, 
or at least metastable, with a sufficiently large lifetime;
therefore it is thus natural to assume that it is
a ground state, or close to,
stationary and rotationally invariant.
Because such a wavefunction is obviously affected by surrounding matter,
parameters $a$'s and $\pwr$ are also environment-dependent.

When it comes to galactic scale effects, this means that only parameters $m$ and $\dnc$  are galaxy-independent;
others would generally vary from galaxy to galaxy,
because every galaxy has its own quantum temperature and localized vacuum wavefunction;
the latter to be regarded as a perturbation of the vacuum wavefunction for a larger superset configuration of gravitating matter.
This resolves the above-mentioned issue of the modified gravity approach.

Furthermore, 
different 
terms of the metric \eqref{e:medsa2} come into play at different length scales.  
The Schwarzschild term, $- r_H/r$, should dominate in the inner regions of galaxies,
where it induces the Keplerian-type orbits.
On the contrary,
the quadratic (de Sitter) term dominates at the largest length scale.
It is probably negligible at a galactic scale, but becomes significant at the scale of clusters and the
observable universe, where it induces an additional mechanism of accelerating expansion;
in addition to that induced by the global flow of background superfluid, 
further details can be found in \Sec 7 of \Ref \cite{z20un1}.
The logarithmic term, $\beta_\pwr^2 
\ln{(\rad/\lnc)}$, predominates in the outer regions of galaxies,
where it induces flat rotation curves.
Interestingly, this term is the only one in \eq \eqref{e:medsa2},
whose behaviour can switch between repulsion and attraction, at the distance $\lnc$.
Finally, the linear term, $r/R_\ssmgl$, becomes significant at extragalactic distances,
but probably at the
outskirts of galaxies as well, depending on a value of $R_\ssmgl$.

These features seem plausible for describing DM- and DE-attributed phenomena \cite{ps20}.
It is also worth noting that while
the linear and quadratic terms were already known to occur in some theories of modified gravity \cite{mk89,gr10,ll13},
the logarithmic term's 
occurrence is, to the best of our knowledge, unique to the logarithmic superfluid vacuum theory.

The
non-relativistic
rotational
velocity curves 
can be estimated using a simple formula
$
v^2 (R)
= 
\frac{1}{2}
r \,
\Der{}{r} 
(
c_{(0)}^2
K^2
)
\bigl|_{r= \ror}
$, 
where 
$\ror$ is the orbit's radius. 
If we take into account the combined contribution 
of gas and stellar disc \cite{too63,cas83},
then we obtain 
\be
v 
=
\sqrt{
v_\text{N}^2 
+ v_\pwr^2 
+
\frac{a_1 b_0}{m \lnc}
R
-
\frac{2 a_2 b_0}{m \lnc^2}
R^2
}
,\lb{e:ov}
\ee
where
\ba
v_\text{N}^2 
&=&
\frac{4}{3}
v_\text{HI}^2
+
v_{\star}^2 
= 
\frac{4}{3}
v_\text{HI}^2
+
\frac{G M_\star}{2 h_R^3} R^2
B\!\left(\frac{R}{2 h_R}\right)
,\lb{e:ovgad}\\
v_\pwr^2 
&=&
\frac{1}{2}
\ror \,
\Der{}{\ror} 
\left[
c_{(0)}^2
\beta_\pwr^2 
\ln{\!\left(
\frac{\ror}{\lnc}
\right)} 
\right]
=
\frac{\pwr b_0}{m} 
, \lb{e:ovgal}
\ea
where
$M_\star$ and $h_R$
are, respectively, the total gravitational mass and surface brightness' scale length of the stellar disc,
$  B (x) = I_0 (x) K_0 (x) - I_1 (x) K_1 (x)$,
$I_{n} (x)$ and $K_{n} (x)$ 
are modified Bessel functions of the first and second kind, respectively, 
and
the mass ratio between helium and neutral hydrogen (HI) is assumed to be $1/3$.
Notice that the contribution \eqref{e:ovgal} is independent of $R$, which
proves our earlier statement about the
relation between the FRC phenomenon and logarithmic term in the induced metric.
Thus, a rotation curve 
would be asymptotically flat if $a_1  = a_2  = 0$ identically, 
while the Keplerian term $v_\text{N}^2$ rapidly decreases as $R$ grows,
thus making the term \eqref{e:ovgal} to predominate in \eq \eqref{e:ov}
at large $R$.
However, if either or both of $a_1$ and $a_2$ 
are not zero then we have what 
we call the \textit{asymptotically non-flat behaviour} of a rotation curve.

Furthermore, let us consider the asymptotic behaviour of orbital velocity
in the utmost outer regions of galaxies, but not much beyond;
therefore we can	 neglect the de Sitter term.
Our goal here is preliminary estimates,
therefore we will not consider gas and disc contributions, for the sake of simplicity
and minimal assumptions for galactic luminous matter.
Fortunately,
because we are considering distances so far away from the galactic center,
almost at the galaxy's border, 
luminous matter contributions should not be of a leading order of magnitude;
although they are still significant \cite{ll13}.

From \eq \eqref{e:ov} we thus obtain
a predicted asymptotic value of this velocity
\be\lb{e:grclin}
v \to
v_\text{(out)}
\approx 
v_\pwr 
\sqrt{
1+
\frac{
\ror}{\lpwr}
},
\ee
where a value
$\lpwr = \pwr \lnc/a_1$
thus
defines the characteristic length scale of the crossover between
flat and linear regimes of rotation curves.
Because $\lnc$ is positive-definite,
the sign of $\lpwr$ equals that of the product of parameters $a_1$ and $\pwr$
of a galaxy's vacuum wavefunction.

\bw\bc
\begin{table}[b]
 \caption{Results of the linear two-parameter best fit to the velocity-squared rotation curves of sample galaxies.}
\bigskip
\begin{tabular}{l|c|c|c|c|c|c}
 \hline
~~~~~~~~~~~~~~~~~~~ &\quad $R_\text{min}$, kpc \quad&\quad $R_\text{max}$, kpc \quad&\quad $\lpwr$, kpc \quad&\quad $v_\text{min}$, km/s \quad&\quad $v_\text{max}$, km/s \quad&\quad $v_\pwr$, km/s \quad\\
\hline
 &  &  &   &  &  &  \\
\quad DDO 154  & 10.96 & 13.78 & 13.47  & 43.62 & 46.97 & 43.81 \\
\quad NGC 2403 & 12.73 & 15.52 & 63.22  & 132.3 & 136.2 & 133.0 \\
\quad NGC 2841 & 34.69 & 44.14 & 389.0  & 251.4 & 264.5 & 260.7 \\ 
\quad NGC 2903 & 22.66 & 28.51 & -91.04 & 178.5 & 182.5 & 181.0 \\
\quad NGC 3198 & 34.15 & 43.18 & 173.0  & 145.6 & 150.3 & 146.4 \\
\quad NGC 3521 & 27.24 & 34.25 & 2439   & 187.0 & 194.0 & 188.1 \\
\quad NGC 5055 & 24.98 & 31.59 & -41.66 & 175.2 & 188.3 & 189.4 \\
\quad NGC 7331 & 29.92 & 39.55 & 159.9  & 221.2 & 238.3 & 233.1 \\
 &  &  &   &  &  &  \\
\hline
\end{tabular}
\lb{t:asd}
\end{table}

\begin{figure}[t]
\centering
\subfloat[DDO154
]{
  \includegraphics[width=\sc\columnwidth]{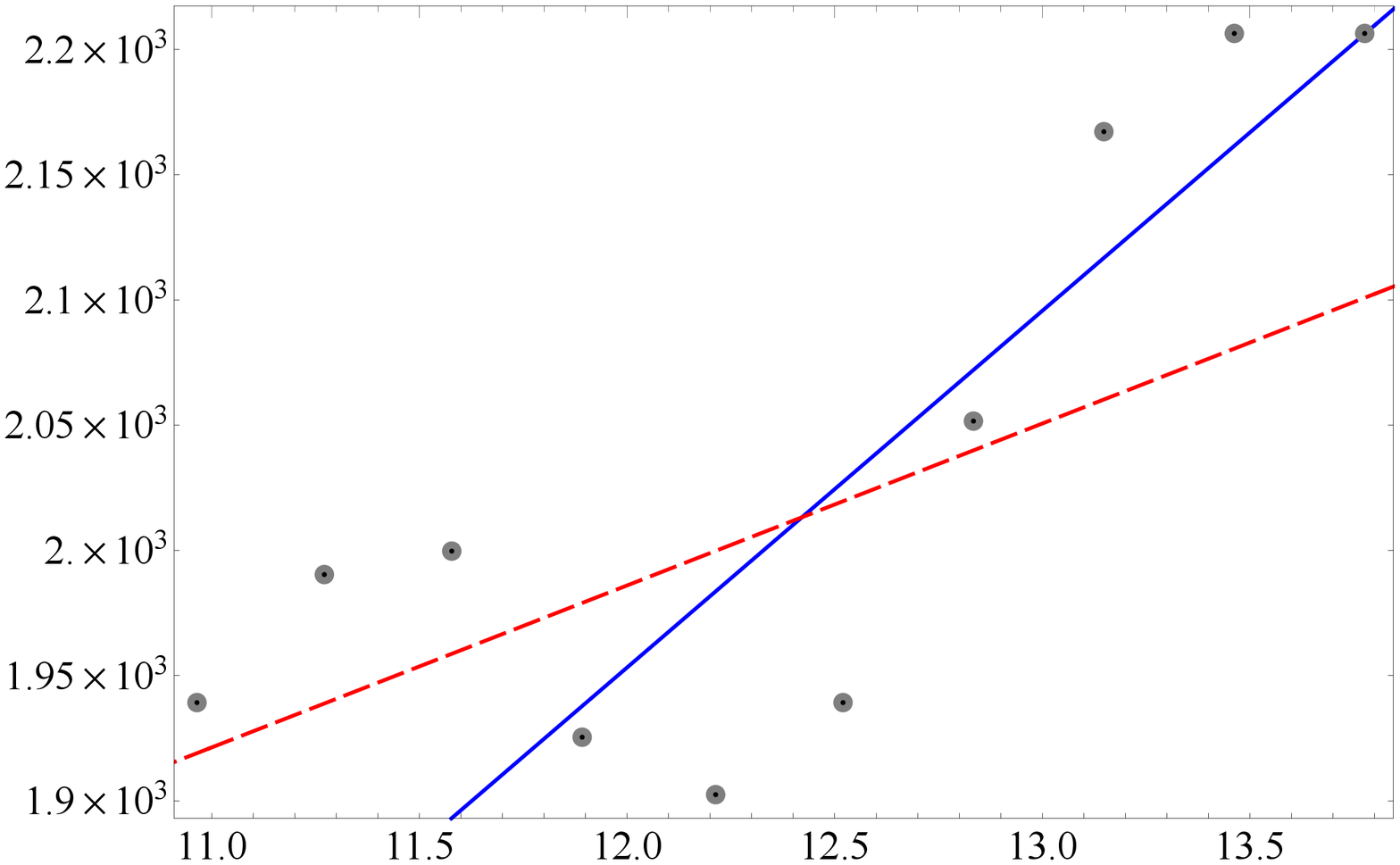}
}
\subfloat[NGC2403
]{
  \includegraphics[width=\sc\columnwidth]{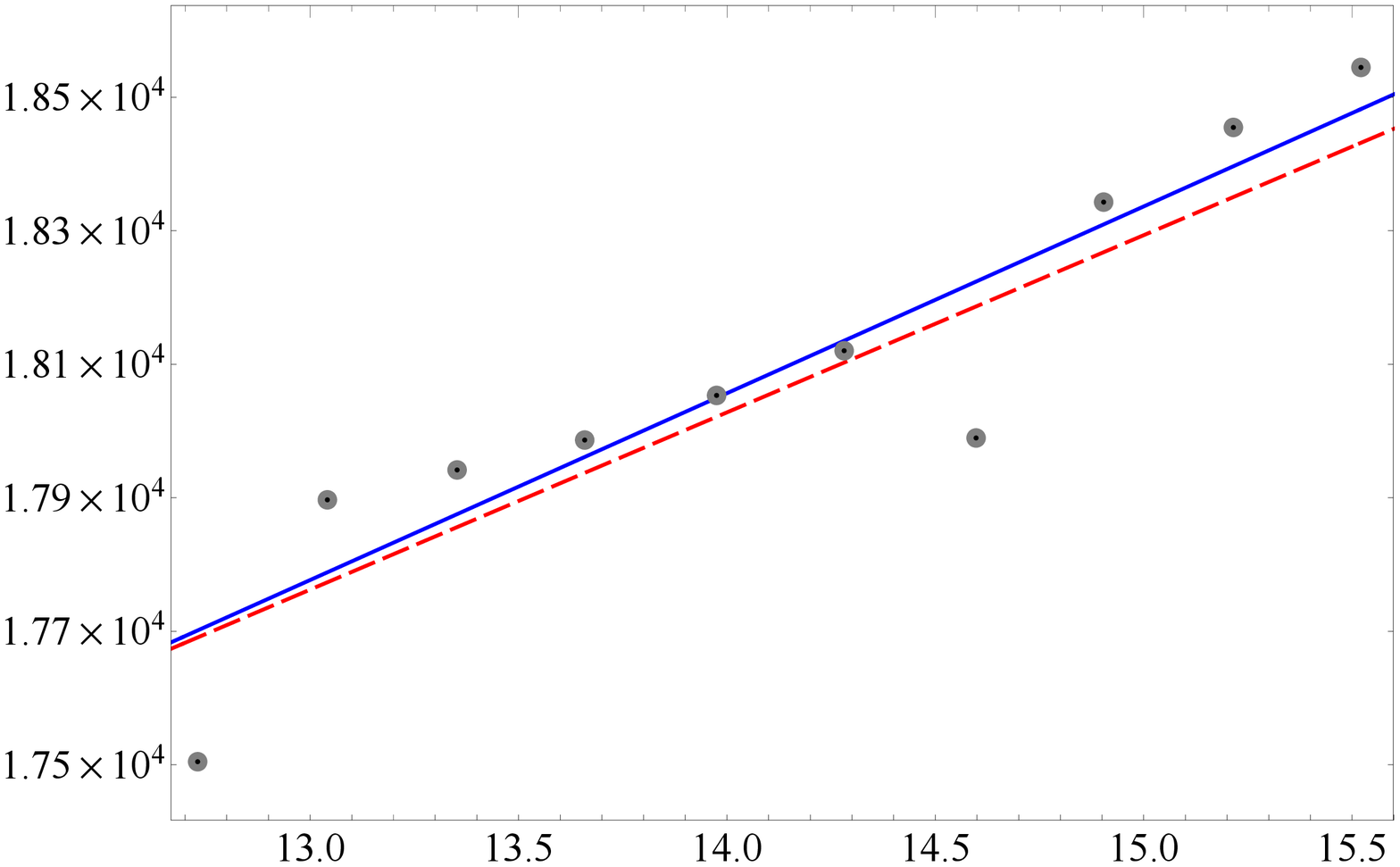}
}
\hspace{0mm}
\subfloat[NGC2841
]{
  \includegraphics[width=\sc\columnwidth]{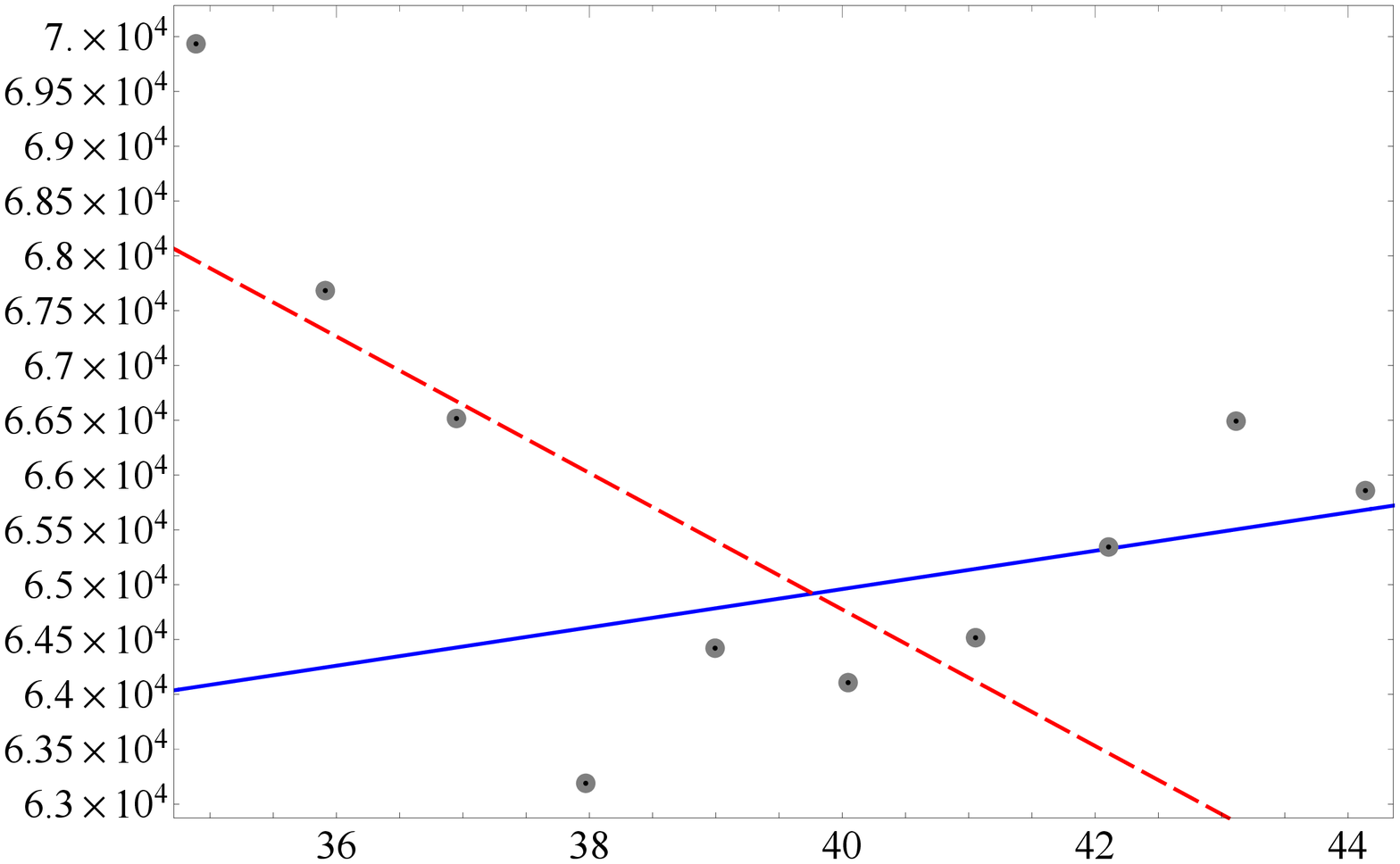}
}
\subfloat[NGC2903
]{
  \includegraphics[width=\sc\columnwidth]{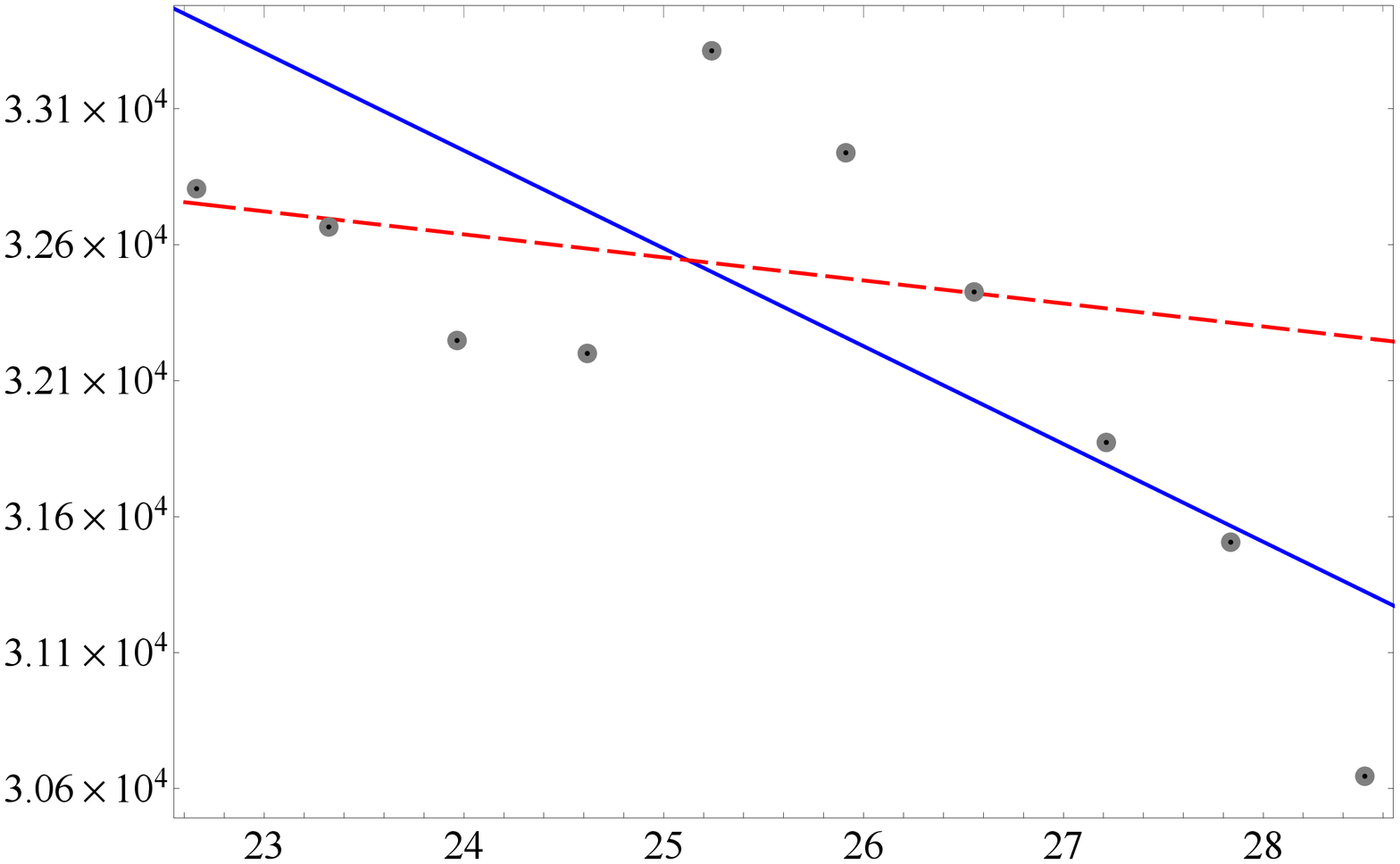}
}
\hspace{0mm}
\subfloat[NGC3198
]{
  \includegraphics[width=\sc\columnwidth]{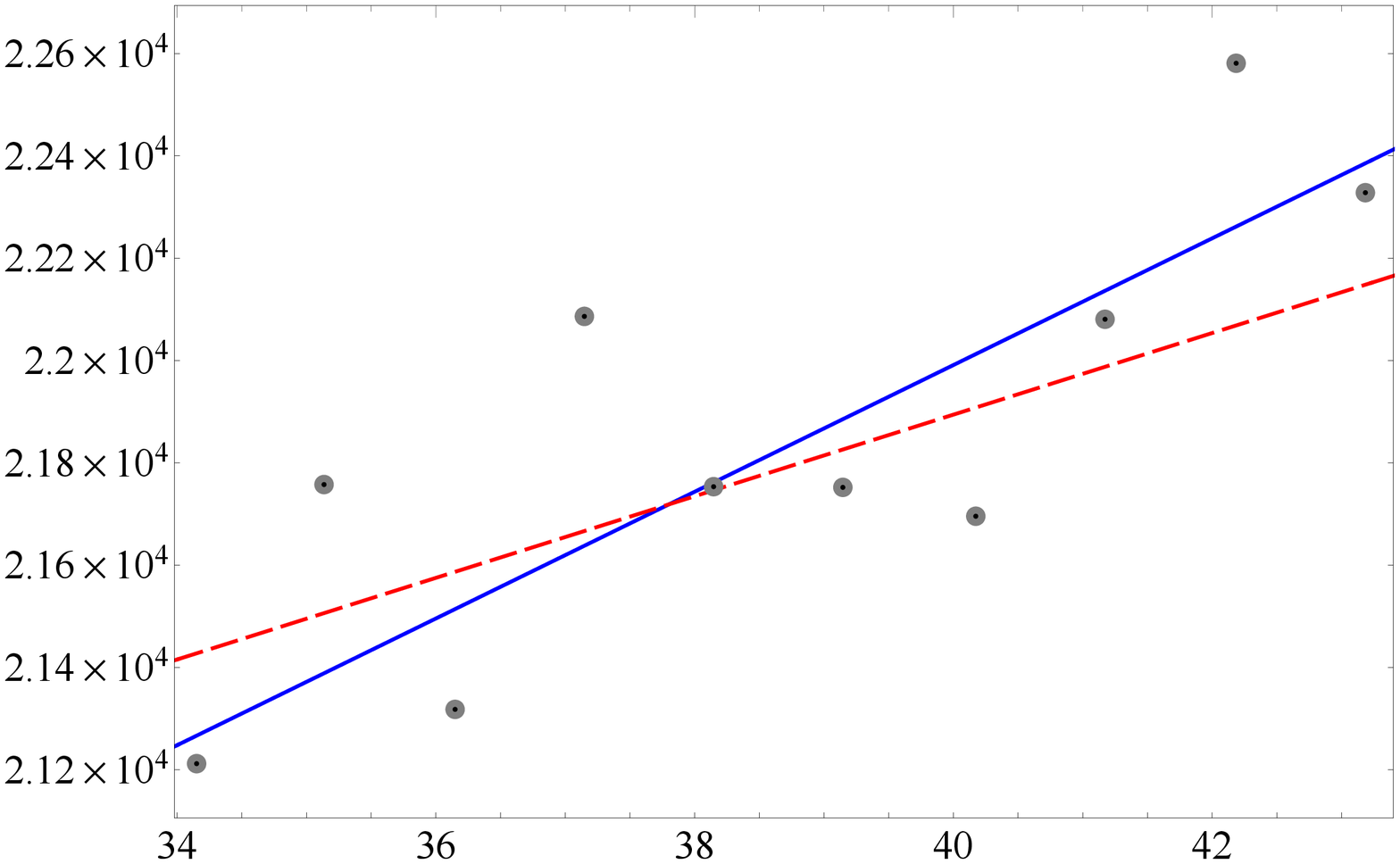}
}
\subfloat[NGC3521
]{
  \includegraphics[width=\sc\columnwidth]{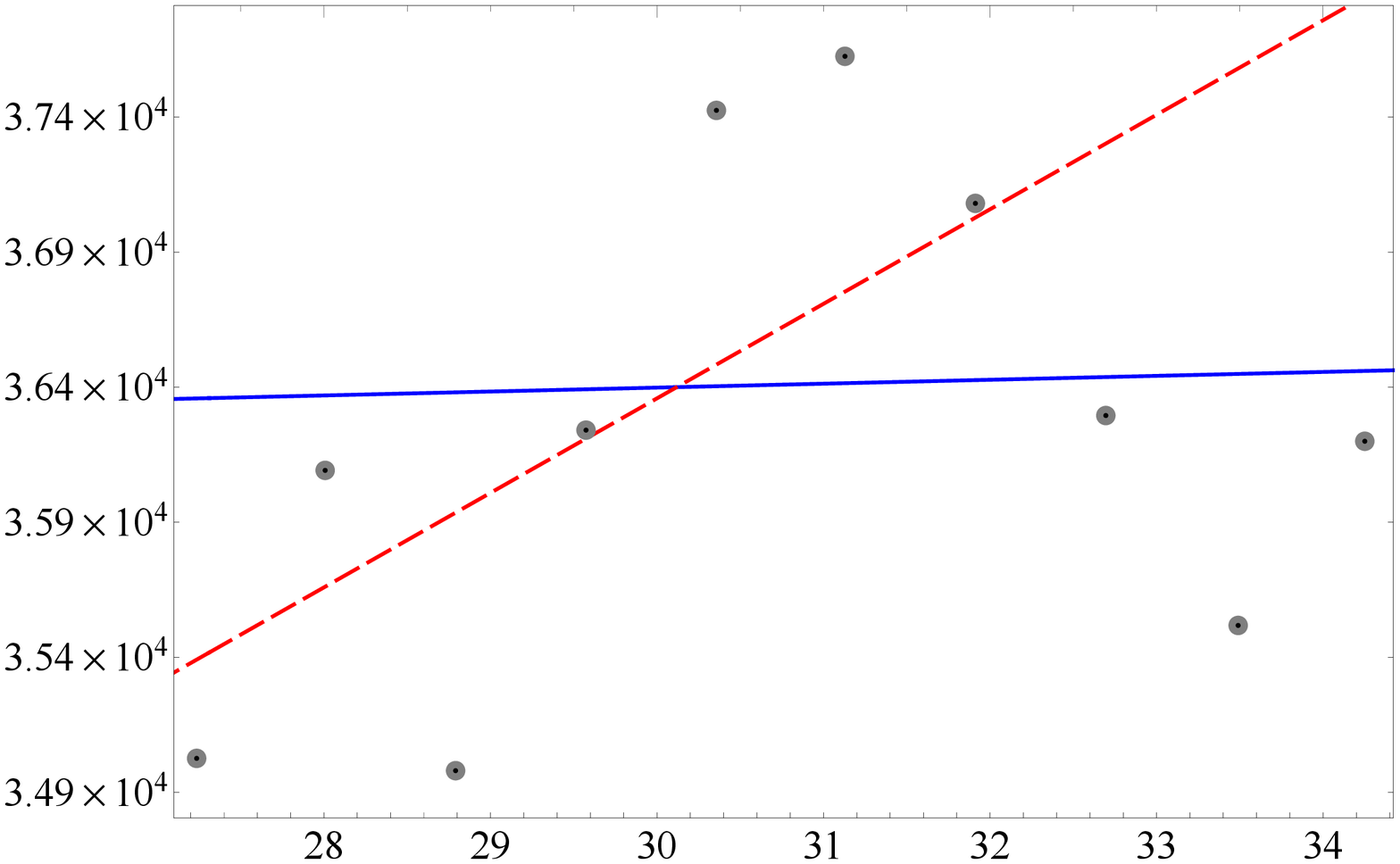}
}
\hspace{0mm}
\subfloat[NGC5055
]{
  \includegraphics[width=\sc\columnwidth]{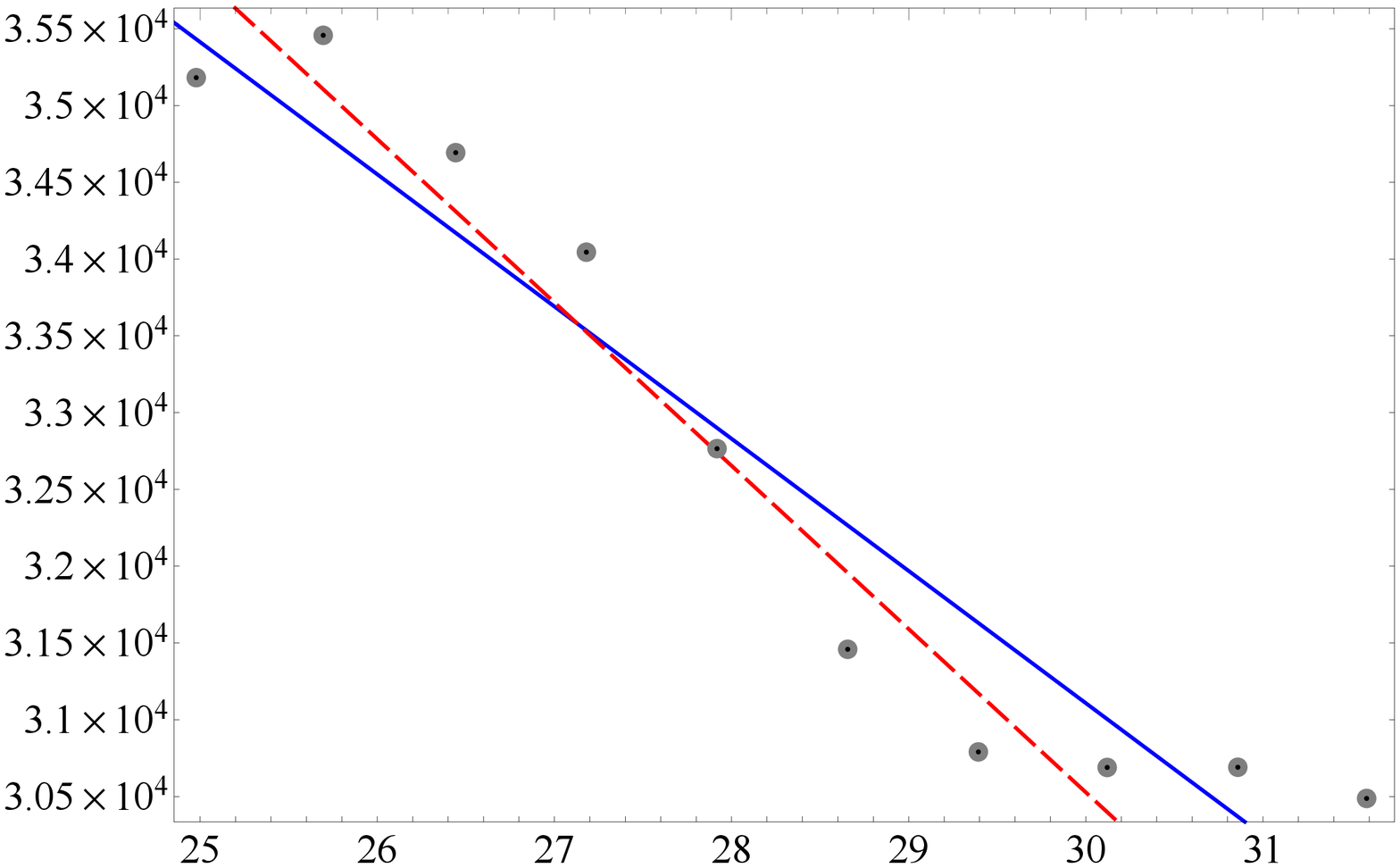}
}
\subfloat[NGC7331
]{
  \includegraphics[width=\sc\columnwidth]{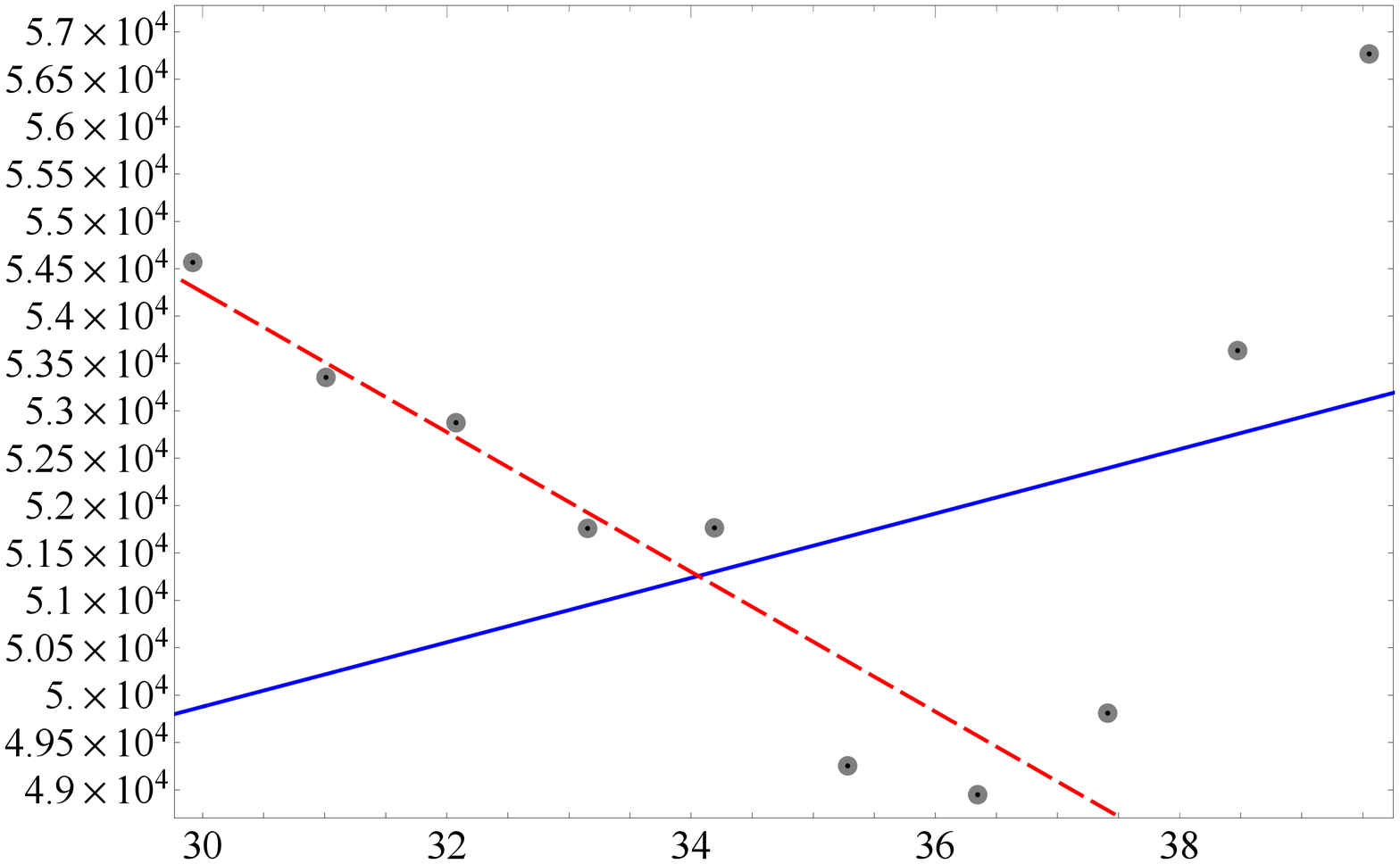}
}
\caption{Linear two-parameter best fit to the velocity-squared rotation curves of sample galaxies.
Black dots refer to mean values of the data points from the THINGS database,
the solid and dashed lines represent the fits of, respectively, main and auxiliary data subsets.
The horizontal axis is $R$ in kpc,
and the vertical axis is $v^2$
in km$^2$/s$^{2}$.
}
\label{f:gal}
\end{figure}
\ec\ew

To analyze experimental data,
it is convenient to square both sides of \eq \eqref{e:grclin} and perform the
least-square fitting of a linear function of radial distance,
$v^2 \mapsto c_1 R + c_0$. 
Sample disc galaxies from the THINGS database are listed in Table \ref{t:asd};
for each of those we took the ten furthest orbits data points known to the database.
From each galaxy's data set we derived two subsets:
the main subset, with the smallest-$R$ twenty per cent of points dropped,
and 
an auxiliary subset, with the highest-$R$ twenty per cent of points omitted.

An 
auxiliary subset is expected to estimate parameters of the flat regime,
such as $v_\pwr $.
This velocity is determined by evaluating the best-fitting line of the subset at the smallest-$R$
point of the subset, $R_\text{min}$.
The data point which corresponds to $R_\text{min}$, is denoted as $v_\text{min}$.

The
main subset is used to estimate possible deviations from the flat regime,
as predicted by \eq  \eqref{e:grclin}.
The data point corresponding to the largest-$R$
point of the subset, $R_\text{max}$, is denoted as $v_\text{max}$.
The value $\lpwr$  is determined as the ratio of the corresponding $v_\pwr^2$ and a linear coefficient
of
the best-fitting line of the subset.

The outcome of the fitting procedure is given in Table \ref{t:asd} and \Fig \ref{f:gal}.
Substantial inclinations of solid lines can clearly be  seen in the figure.
This indicates the presence of a crossover transition from  flat to linear regimes of rotation
with increasing $R$,
as predicted by \eq \eqref{e:grclin}:
because tangent lines of best fits become no longer exactly horizontal
for the furthest data points,
then
rotation curves become non-flat as one approaches galactic boundaries.

For a number of galaxies, NGC 2841, NGC 3198, NGC 7331 and especially NGC 3521,
the computed characteristic linear-regime length value $|\lpwr|$ turns out 
to be much larger than $ R_\text{max} $, by one, or even two orders of magnitude.
This is because the majority of data points still belong to the flat regime; 
while the full crossover to the post-flat regime occurs beyond the chosen data range.
This is still compatible with the theory;
because the linear term in the metric \eqref{e:medsa2} is expected to
predominate at extragalactic length scales,
while the logarithmic term is most significant in the outer regions of galaxies. 

These preliminary estimates are sufficient for the purposes of the current \art,
while
the detailed fitting of rotation curves, which obviously require 
the combined contribution \eqref{e:ovgad} from the gas and stellar disc been taken into account,
will remain as the subject of future studies.


\section*{Funding}
This work is based on the research supported by Department of Higher Education and Training of South Africa, and in part by the National Research Foundation of South Africa
(Grants Nos. 95965, 132202 and 131604).

\begin{acknowledgments}
Proofreading of the manuscript by P. Stannard is greatly appreciated.
\end{acknowledgments}













\end{document}